\documentclass[aps,prd,amsmath,amssymb,nofootinbib,preprintnumbers,twocolumn]{revtex4-1}
 
\usepackage{hyperref} 
\usepackage{graphicx,xcolor}% Include figure files 
\usepackage{epstopdf} 
\usepackage{xcolor}

\def\pa{{\partial}}
\newcommand{\be}{\begin{equation}}
\newcommand{\ee}{\end{equation}}
\newcommand{\ba}{\begin{eqnarray}}
\newcommand{\ea}{\end{eqnarray}}
\newcommand{\nn}{\nonumber}

\newcommand{\h}[1]{\hat{#1}}

\newcommand{\hook}{\raisebox{-0.35ex}{\makebox[0.6em][r]
{\scriptsize $-$}}\hspace{-0.15em}\raisebox{0.25ex}{\makebox[0.4em][l]{\tiny
 $|$}}}

\begin{document} 
\title{General theory of Galilean gravity}

\author{Marco Cariglia}
\email{Marco.Cariglia@ufop.edu.br}
\affiliation{Departamento de F\'isica, Universidade Federal de Ouro Preto, 35400-000 Ouro Preto MG, Brazil}

%\date{\today}  % version

\begin{abstract} 
We obtain the complete theory of Newton-Cartan gravity in a curved spacetime by considering the large $c$ limit of the vielbein formulation of General Relativity. Milne boosts originate from local Lorentzian transformations, and the special cases of torsionless and twistless torsional geometries are explained in the context of the larger locally Lorentzian theory. We write the action for Newton-Cartan fields in the first order Palatini formalism, and the large $c$ limit of the Einstein equations.  Finally, we obtain the generalised Eisenhart-Duval lift of the metric that plays an important role in non-relativistic holography. 
\\ 

\vspace{0\baselineskip} \noindent\textbf{Keywords}: 
Netwon-Cartan structures with torsion, non-relativistic AdS/CFT, Eisenhart-Duval lift. \\ 
 
\begin{center} \textit{In memoriam Christian Duval} \end{center} 
\end{abstract}  
 
\date{\today}

\maketitle 
As the concept of AdS/CFT correspondence \cite{maldacena1999large,gubser1998gauge,witten1998anti} is maturing, the focus of its applications has gradually shifted from fundamental field theories to condensed matter systems, see for example \cite{hartnoll2009lectures,hartnoll2008building,gubser2008gravity}. In the frame of the latter AdS/CMat correspondence there has been lot of activity in recent years in understanding and setting up a non-relativistic version of holography  \cite{SonWingate2006,Son2008,BalasubMcGreevy2008,HotSpacetimes2008,HerzogEtAl2008,Son2013,SonEtAl2014,maldacena2008}. This is related with the notion of Galilei-invariant field theories in curved spacetime \cite{Jensen2014,Banerjee:2014pya,Banerjee:2014nja,Banerjee:2015rca,Banerjee:2017rch,Banerjee:2018gqz}, and the concept of Newton-Cartan (NC) structures plays an important role \cite{Kunzle1972,ChristianKunzle1984,DBKP1985,Christian1993,ChristianPeter2009,ChristianGaryPeter1991}. In fact, gauging the Bargmann algebra, i.e., the centrally extended Galilean algebra, yields a torsionless Newton-Cartan structure \cite{BergshoeffEtAl2013}, and the same procedure allows defining a Newton-Cartan supergravity \cite{andringa20133d}. It has been argued that in the context of holography the appropriate Newton-Cartan structure should include torsion: this arises in specific holographic models that can be embedded in String Theory \cite{HartongEtAl2014,HartongEtAl2014_2}, in formulations where one gauges the Schr\"{o}dinger algebra \cite{BergshoeffEtAl2015,bergshoeff2015newton}, in Schr\"{o}dinger holography \cite{hartong2015lifshitz}, and in studies of non-relativistic physical systems made using geometrical  methods\cite{gromov2015thermal,geracie2015hydrodynamics}. Torsional Newton-Cartan structures (TNC) are furthermore divided into the simpler twistless torsional structures (TTNC), where the `clock' vector $n_\mu$ is hypersurface orthogonal, thus allowing the definition of a notion of time, and the more general uncontrained case. The TTNC is better understood, quite recently it has been shown how it arises from the large $c$ expansion of General Relativity \cite{van2017torsional} in the absence of matter, and how to build an action principle for it \cite{hansen2018action}. The generic torsional case is not as clear.  

% which can be seen to arise as a non-relativistic limit of General Relativity\cite{BergshoeffEtAl2015_2}.  

In this work we present the complete theory of torsional Newton-Cartan gravity, without assuming that the torsion is twistless. We do this by studying the large $c$ expansion of General Relativity in the vielbein formalism. This allows us to produce an action for Newton-Cartan fields that is obtained from the Einstein-Hilbert action in the first order Palatini formalism, as well as displaying the full equations of motion for the fields, by taking the large $c$ limit of the Einstein field equations. By studying the decomposition of the spin connection we can provide a clear geometrical description of the conditions associated to zero torsion and twistless torsion. Milne boosts are seen to arise from the large $c$ limit of Lorentz boosts, thus justifying their identification in the literature with Galilean boosts. We remark that Newton-Cartan theories with unconstrained torsion do not admit the definition of a time variable. This is a radical departure from the non-relativistic theories of gravity known so far. The physical implications of such absence of time probably still need to be appropriately understood. We take the pragmatic approach that pursuing an understanding of such theories is useful for two main reasons: on one side the applications to non-relativistic holography, as mentioned before, and on the other the fact that such theories arise in the appropriate limit of General Relativity in the presence of matter, thus opening a different angle to study these latter important systems. 
 
Our initial motivation for this research was to have a deeper understanding of the torsion condition. In non-relativistic holography the bulk metrics are a pp-wave generalization of AdS spaces \cite{duval2009geometry}, clearly related to so called Eisenhart-Duval lift metrics \cite{Eisenhart1928,DBKP1985,ChristianGaryPeter1991,CarigliaRMP2014,Galajinsky2016,TimeDependent2016,cariglia2018cosmological}. However, Eisenhart-Duval lift metrics have always been studied in the context of zero torsion. It was then natural to enquire about the meaning of torsion. Since our enquiry was of a geometrical nature, we started from the beginning with a decomposition the vielbeins of General Relativity. This lead us to a further decomposition of the remaining fields, the spin connection and the Riemann tensor. From the general relativistic metric we then built a higher dimensional generalized Eisenhart-Duval lift metric that includes torsion and describes the dynamics of non-relativistic particles in the original curved spacetime. It then becomes apparent that torsion models the non-trivial geometry in which the particles move. We expect this metric, through its Fefferman-Graham expansion, to be relevant for the non-relativistic holography bulk dynamics. 
 
The general theory described in this letter will be useful in three separate settings. First, for studying strong gravitational fields in a non-relativistic setting in the presence of matter, second for the study of non-relativistic holography, and third for generalizing the concept of Eisenhart-Duval lift in the framework of dynamical systems. 
 
When this work was substantially complete, we learned about the results in \cite{hansen2018action}, which overlap with ours. In our case the TTNC constraint is removed.  
 
\noindent \textbf{A brief review of Newton-Cartan geometries} \\ 
A Newton-Cartan structure on a non-relativistic $d+1$-dimensional spacetime $M$ is defined by the following set of geometrical objects: $\{ n_\mu, h^{\mu\nu}, v^\mu, A_\mu \}$. $n = n_\mu \, dx^\mu$ is a nowhere-vanishing 1--form, the clock, that defines locally a time direction. $h^{\mu\nu}$ is a simmetric tensor that is semi-positive-definite, of rank $d$, and whose kernel is generated by $n$, $h^{\mu\nu} n_\nu = 0$. One can think of the subspace on which $h^{\mu\nu}$ is positive definite as the 'spatial slice' of spacetime at a certain point in $M$. In order to be able to raise and lower indices as follows one first chooses a vector $v^\mu$ such that $n_\mu v^\mu = 1$. 
Then $v^\mu$ must also be nowhere-vanishing. Such a vector is not unique: any other vector $v^{\prime\mu}$ that satisfies the same condition can be written as 
\be \label{eq:Milne_v}
v^{\prime\mu} = v^\mu + h^{\mu\nu} \psi_\nu \, ,  
\ee 
for a generic $\psi_\mu$. In the literature this is called a \textit{Milne boost}, and it is associated to the Galilean boosts in the Schr\"{o}dinger algebra. Given $v^\mu$ one can now define a new tensor $h_{\mu\nu}$ that satisfies the two conditions 
\be 
v^\mu h_{\mu\nu} = 0\, , \qquad h_{\mu\nu} h^{\nu\rho} = \delta^\rho_\mu - n_\nu v^\rho =: P^\rho_\nu \, . 
\ee 
This defines a projector that projects on spatial slices, and $h_{\mu\nu}$, $h^{\mu\nu}$ are mutually inverse on the subspace of $TM$ that is in the image of $P$. The ambiguity in the choice of $v^\mu$ is reflected in a related ambiguity in the choice of $h_{\mu\nu}$: if $v^\mu$ changes by \eqref{eq:Milne_v} then $h_{\mu\nu}$ has to change according to 
\be \label{eq:Milne_h}
h^\prime_{\mu\nu} = h_{\mu\nu} - \left( n_\mu P_\nu^\rho + n_\nu P_\mu^\rho \right) \psi_\rho + n_\mu n_\nu h^{\rho\sigma} \psi_\rho \psi_\sigma \, . 
\ee 
 
The form $A_\mu$ included in the definition of a Newton-Cartan structure appers when defining covariant derivatives, which we indicate with the symbol $D$, while we indicate with $\gamma$ the Newton-Cartan connection. It is customary to ask that the covariant derivative satisfies the following two compatibility conditions: 
\be 
D n_\mu = 0\, , \quad D h^{\mu\nu} = 0 \, . 
\ee
In \cite{Jensen2014}  it is shown that these  conditions, plus the request that the torsion has a purely temporal upper index, i.e. $h_{\lambda\rho} \left( \gamma^\lambda_{\mu\nu} - \gamma^\lambda_{\nu\mu} \right) = 0$, fixes the form of the connection to be 
\ba \label{eq:NC_connection} 
\gamma^\lambda_{\mu\nu} &=& v^\lambda \pa_\mu n_\nu + \frac{1}{2} h^{\lambda\rho} \left( \pa_\mu h_{\rho\nu} + \pa_\nu h_{\mu\rho} - \pa_\rho h_{\mu\nu} \right) \nn \\ 
&& + h^{\lambda\rho} n_{(\mu} F_{\nu) \rho} \, .  
\ea 
The torsion is given by $v^\lambda \left( \pa_\mu n_\nu - \pa_\nu n_\mu \right) = v^\lambda dn_{\mu\nu}$, while the last term in \eqref{eq:NC_connection} is tensorial and can be chosen at will for any completely antisymmetric 2--form $F_{\mu\nu} = F_{[\mu\nu]}$. The connection has been  called \textit{Newtonian} if $dF = 0$, i.e. $F = dA$ for a vector potential $A_\mu$ which has often been interpreted as the gauge field associated to the conservation of mass, or particle number. When $dn=0$ is it possible to show that the connection \eqref{eq:NC_connection} is left invariant under a Milne boost if $A_\mu$ also changes as 
\be \label{eq:Milne_A} 
A^\prime_\mu = A_\mu + P_\mu^\nu \psi_\nu - \frac{1}{2} n_\mu h^{\rho\sigma} \psi_\rho \psi_\sigma \, . 
\ee 
When $dn\neq 0$ instead one finds that it is not possible to have a connection that is Milne invariant and at the same time $U(1)$ invariant. The most frequently taken path consists in keeping the transformation law \eqref{eq:Milne_A} and redefining the connection by the following tensorial part 
\ba \label{eq:Gamma_Milne_invariant}
\tilde{\gamma}{}^\lambda_{\mu\nu} &=& \gamma^\lambda_{\mu\nu} + h^{\lambda\sigma} \left( A_\mu \pa_{[\nu} n_{\sigma]} + A_\nu \pa_{[\mu} n_{\sigma]} - A_\sigma \pa_{[\mu} n_{\nu]} \right) \nn \\ 
&=& \tilde{v}^\lambda \pa_\mu n_\nu \hspace{-0.05cm} + \hspace{-0.05cm} \frac{1}{2} h^{\lambda\rho} \hspace{-0.05cm} \left( \pa_\mu \tilde{h}_{\rho\nu} + \pa_\nu \tilde{h}_{\mu\rho} - \pa_\rho \tilde{h}_{\mu\nu} \right)  , 
\ea 
where $\tilde{v}^\mu = v^\mu - h^{\mu\nu} A_\nu$, $\tilde{h}_{\mu\nu} = h_{\mu\nu} + n_\mu A_\nu + n_\nu A_\mu$ are Milne invariant. The resulting connection $\tilde{\gamma}$ is Milne invariant but is not $U(1)$ invariant. The new fields satisfy 
\be 
\tilde{h}_{\mu\nu} \tilde{v}^\nu = s n_\mu\, , \qquad h^{\sigma\nu} = \delta_\mu^\nu - n_\mu \tilde{v}^\nu \, , 
\ee 
where $s = 2 A_\mu v^\mu - h^{\mu\nu} A_\mu A_\nu$ is a scalar that is Milne invariant but not $U(1)$ invariant. The new covariant derivative $\tilde{D}$ obtained from $\tilde{\gamma}$ preserves the previous compatibility conditions: $\tilde{D} n_ \mu = 0$, $\tilde{D} h^{\mu\nu} = 0$. The geometry is called torsionless when $dn=0$, and twistless torsional when $n \wedge dn = 0$, i.e. $n$ is hypersurface orthogonal. \\ 
\textbf{Expansion of the vielbeins} \\ 
We start with the spacetime of General Relativity, modelled with a spin manifold $M$ of dimension $d+1$:  coordinates on $M$ are given by $x^\mu$, $\mu = 0, 1, \dots, d$. Vielbeins are forms $e^{\h{a}}_\mu \, dx^\mu$, where $\h{a}=0,1,\dots , d$ are orthonormal indices. We will use the split $\h{a} = (\h{0}, \h{i})$, with $\h{i}=1, \dots, d$. Inverse vielbeins are denoted by $E^\mu_{\h{a}}$ and satisfy $e_\mu^{\h{a}} E^\mu_{\h{b}} = \delta^{\h{a}}_{\h{b}}$, $e^{\h{a}}_\mu E^\nu_{\h{a}} = \delta^\nu_\mu$. The general relativistic metric on $M$ will be denoted by $g_{\mu\nu}$ and $e^{\h{a}}_\mu e^{\h{b}}_\nu \eta_{\h{a}\h{b}} = g_{\mu\nu}$, where $\eta_{\h{a}\h{b}} = \text{diag} (-c^2, 1, \dots, 1)$, and similarly $E^\mu_{\h{a}} E^\nu_{\h{b}} g_{\mu\nu} = \eta_{\h{a}\h{b}}$. Therefore the dimensionality of the inverse vielbeins is $\left[ E_{\h{i}}^\mu \right] = 1$, $\left[ E_{\h{0}}^\mu \right] = L T^{-1}$, while for vielbeins $\left[ e^{\h{i}}_\mu \right] = 1$, $\left[ e^{\h{0}}_\mu \right] = L^{-1} T$. Our starting premise is the parameterization of the vielbeins in the large $c$ limit 
\ba 
e^{\h{0}}_\mu &=& n_\mu - \frac{1}{c^2} A_\mu + O(c^{-4})\, , \\ 
e^{\h{i}}_\mu &=& l^i_\mu + \frac{1}{c^2} m^i_0 n_\mu + \frac{1}{c^2} m^i{}_j l^j_\mu + O(c^{-4}) \, , \label{eq:ei_expansion} \\ 
E_{\h{0}}^\mu &=& \left(1 + \frac{1}{c^2} A_0 \right) v^\mu - \frac{1}{c^2} m^i_0 L_i^\mu + O(c^{-4}) \, , \\ 
E_{\h{i}}^\mu &=& L_i{}^\mu + \frac{1}{c^2} A_i v^\mu - \frac{1}{c^2} m^j{}_i L_j^\mu + O(c^{-4}) \, .  
\ea  
Here $m^i{}_\mu$ is a 1--form and we used the notation $A_0 = v^\mu A_\mu$, $m^i_0 = v^\mu m^i{}_\mu$, $m^i{}_j = L_j^\mu m^i{}_\mu$. Notice how we have introduced two different projections for the indices of a generic tensor $T^{\lambda_1 \dots \lambda_p}_{\mu_1 \dots \mu_q}$: 
\ba 
T^{\h{0} \dots \h{l}_p}_{\h{0} \dots \h{m}_q} &=& e^{\h{0}}_{\lambda_1} \dots e^{\h{l}_p}_{\lambda_p} E_{\h{0}}^{\mu_1} \dots E_{\h{m}_q}^{\mu_q} \, \, T^{\lambda_1 \dots \lambda_p}_{\mu_1 \dots \mu_q} \quad \text{vielbein,} \nn \\ 
T^{0 \dots l_p}_{0 \dots m_q} &=& n_{\lambda_1} \dots l^{l_p}_{\lambda_p} v^{\mu_1} \dots L_{m_q}^{\mu_q} \, \, T^{\lambda_1 \dots \lambda_p}_{\mu_1 \dots \mu_q} \quad \text{Newton-Cartan.} \nn 
\ea

Expanding the orthogonality conditions for the vielbeins we find 
\be 
n_\mu v^\mu = 1\, , \quad n_\mu L^\mu_i = 0 \, , \quad  v^\mu l^i_\mu=0\, , \quad l^i_\mu L^\mu_j = \delta^i_j\, . 
\ee 
Therefore setting $h_{\mu\nu} := l^i_\mu l^i_\nu$, $h^{\mu\nu} := L_i^\mu L_i^\nu$, then $\{n_\mu, v^\mu, h_{\mu\nu}, h^{\mu\nu} \}$ satisfy the algebraic conditions for a Newton-Cartan structure. The metric and inverse metric are expanded as 
\ba \label{eq:metric_expansion}
\hspace{-0.35cm} g_{\mu\nu} =  - c^2 n_\mu n_\nu + \left( h_{\mu\nu} + 2 n_{(\mu} A_{\nu)} \right) + O\left( c^{-2} \right) \, , 
\ea 	
\ba \label{eq:inverse_metric_expansion}
g^{\mu\nu} &=&  h^{\mu\nu} - \frac{1}{c^2}  \left( v^\mu v^\nu - 2 h^{(\mu| \rho} A_\rho v^{\nu)}  - \beta^{\mu\nu}  \right) \nn \\ 
&& + O\left(c^{-4} \right) \, , 
\ea 
where $\beta^{\mu\nu} = - 2 L_i^{(\mu|} m^j{}_i L_j^{\nu)}$. 
 
General Relativity is a gauge theory of local Lorentz transformations. A generic Lorentz boost can be written, modulo a rotation, as $t^\prime = \gamma \left( t - \frac{\beta}{c} x_1 \right)$, $x^\prime = \gamma \left( x - u t \right)$, where $u \in \mathbb{R}$ is the speed parameter, $\beta = \frac{u}{c}$, $\gamma = (1 - \beta^2)^{-\frac{1}{2}}$. It is well known that in the limit $u \rightarrow + \infty$ these reduce to Galilean boosts $t^\prime = t$, $x^\prime =  x - u t$. This is encoded in our vielbein expansion: acting on $e^{\h{0}}_\mu$ with a Lorentz boost and expanding to order $c^{-2}$ we obtain 
\ba 
n^\prime_\mu &=& n_\mu\, , \\ 
A^\prime_\mu &=&  A_\mu + u \alpha_i l^i_\mu - \frac{u^2}{2} n_\mu\, , 
\ea 
where $\alpha_i$ is of modulus one and determines the spatial direction of the Lorentz boost. If we write $u\alpha_\mu = \psi_\mu$, then this is the transformation law \eqref{eq:Milne_A}. Next, Lorentz boosting $e^{\h{i}}_\mu$ we find 
\be 
l^{\prime i}_\mu = l^i_\mu - \psi_i n_\mu \, , 
\ee 
which implies \eqref{eq:Milne_h}. 
The metric is boost invariant under these transformations. Boosting $E^\mu_{\h{0}}$ we obtain eq.\eqref{eq:Milne_v}, and finally boosting $E^{\h{i}}_\mu$ we find that $L_i^\mu$ is invariant, 
which implies the Milne invariance of $h^{\mu\nu}$. With these transformations it can be confirmed that the inverse metric is also boost invariant, as it should be by construction. Therefore standard Milne transformations arise from the large $c $ limit of Lorentz boosts. 

The Milne invariant objects $\tilde{h}_{\mu\nu}$ and $\tilde{v}^\mu$  in our formalism are given by $\tilde{h}_{\mu\nu} = g_{(0)\mu\nu}$, which inherits Milne invariance from the local Lorentz invariance of $g_{\mu\nu}$, and similarly $\tilde{v}^\mu = - g^{(0)\mu\nu} n_\nu$. The Levi-Civita connection of $g_{\mu\nu}$, which we indicate with $\Gamma^\lambda_{\mu\nu}$, can also be expanded. We find 
\ba \label{eq:Gamma_GR_2}
\Gamma^\lambda_{\mu\nu} &=&  c^2 h^{\lambda\rho} n_{(\mu} dn_{|\rho|\nu)} +  \tilde{\gamma}^\lambda_{\mu\nu} + W^\lambda_{\mu\nu} \, , 
\ea 
where  
\ba 
W^\lambda_{\mu\nu} &=&  \left(  v^\lambda v^\rho - 2 h^{(\lambda| \sigma} A_\sigma v^{\rho)}  + 2 L_i^{(\lambda|} m^j{}_i L_j^{\rho)}  \right) \nn \\ 
&& \hspace{1cm} \left( n_\mu \pa_{[\nu} n_{\rho]} +n_\nu \pa_{[\mu} n_{\rho]} - n_\rho \pa_{[\mu} n_{\nu]} \right) 
\ea 
is the same tensor that appears in \cite{van2017torsional}. Going forward we will decompose Levi-Civita covariant derivatives in terms of Milne invariant ones using \eqref{eq:Gamma_GR_2}.

\noindent \textbf{Expansion of the spin connection and action for the NC fields}. \\ 
We will assume that the general relativistic manifold has no torsion. If $\nabla$ is the Levi-Civita covariant derivative, then we can introduce a new covariant derivative $\hat{\nabla}$ on objects with mixed curved and orthonormal indices in the standard way asking that $\hat{\nabla}_\lambda e^{\h{a}}_\mu := \nabla_\lambda e^{\h{a}}_\mu + \omega_\lambda{}^{\h{a}}{}_{\h{b}} \, e^{\h{b}}_\mu = 0$. $\omega_\lambda{}^{\h{a}}{}_{\h{b}}$ is the spin connection that can be expressed as 
\be 
\omega_\lambda{}^{\h{a}}{}_{\h{b}} = - E^\mu_{\h{b}} \, \nabla_\lambda e_\mu^{\h{a}} \, . 
\ee 
Using \eqref{eq:Gamma_GR_2} we find that that $\omega_\mu{}^{\h{0}}{}_{\h{j}}$ has no term proportional to $c^2$. This is is related to the fact that it must be $\omega_\mu{}^{\h{0}}{}_{\h{j}} = \frac{1}{c^2} \omega_\mu{}^{\h{j}}{}_{\h{0}}$. 
We write the expansion as 
\be 
\omega_\mu{}^{\h{0}}{}_{\h{j}} =  \overset{\scriptscriptstyle(0)}{\omega}_{\mu}{}^{\h{0}}{}_{\h{j}}  + \frac{1}{c^2}  \overset{\scriptscriptstyle(-2)}\omega_{\mu{}}{}^{\h{0}}{}_{\h{j}}  + O(c^{-4}) \, . 
\ee 
At order $c^0$ 
\be \label{eq:omega0j_zero}
 \overset{\scriptscriptstyle(0)}\omega{}^{\h{0}}{}_{\h{j}}  =  \frac{1}{2} L_j \hook \left( dn + n \wedge ( v \hook dn) \right) \, . 
\ee 
This is not Milne invariant, because the spin connection will transform  under a Lorentz transformation. From eq.\eqref{eq:omega0j_zero}  $\overset{\scriptscriptstyle(0)}{\omega}{}^{\h{0}}{}_{\h{j}}  = 0$ if and only if $dn = 0$, thus  providing a geometrical interpretation within GR of the torsionless property of Newton-Cartan geometry. This is obvious when $dn = 0$. To see that the converse is  true, we parameterize 
\be \label{eq:dn_expansion}
dn = n \wedge (\rho_i l^i) + \frac{1}{2} \sigma_{ij} l^i \wedge l^j = n \wedge \rho + \sigma \, ,  
\ee 
with $v^\mu \rho_\mu = 0$, $v^\mu \sigma_{\mu\nu} = 0$. 
Although the individual $l^i$ are not Milne invariant, one can see that the combination in $dn$ is Milne invariant, as it should be, if under a Milne transformation 
\ba 
\rho_i^\prime &=& \psi_j \sigma_{ji} \, , \\ 
\sigma_{ij}^\prime &=& \sigma_{ij} \, . 
\ea 
With this parameterisation $ \overset{\scriptscriptstyle(0)}\omega{}^{\h{0}}{}_{\h{j}}  = -  \rho_j n + \frac{1}{2} \sigma_{jk} l^k$. Then this is zero if and only if $\rho = 0$, $\sigma = 0$. The statement $ \overset{\scriptscriptstyle(0)}\omega{}^{\h{0}}{}_{\h{j}}  = 0$ is Milne invariant, so it will be true at order $c^0$ in any vielbein frame of the general relativistic theory, modulo global Lorentz transformations. TTNC geometries instead are those with $\sigma = 0$, and are characterised by $ e^{\h{0}} \wedge  \omega^{\h{0}}{}_{\h{j}}  = 0$ at order $c^0$. 
The next term in the expansion of $\omega^{\h{0}}{}_{\h{j}}$ is 
\ba 
&& \overset{\scriptscriptstyle(-2)}\omega_{\mu}{}^{\h{0}}{}_{\h{j}}  = L_j^\rho \tilde{D}_{\mu} A_{\rho} + \left( \frac{1}{2}\sigma_{\mu k} m^k{}_j + \frac{1}{2} \left( A_0 - A_\perp^2  \right)  \sigma_{\mu j}  \right) \nn \\ 
&& + n_\mu \left( \rho_k m^k{}_j + (A_0  - A_\perp^2) \rho_j - \frac{1}{2} A_\nu \beta^{\nu\rho} \sigma_{\rho j} - \frac{1}{2} A_0 A_k \sigma_{kj}   \right) \nn \\ 
&& + r_\mu{}^0{}_j \, ,  
\ea 
where $A_\perp^2 = h^{\mu\nu} A_\mu A_\nu$. The form $r^0{}_j$ is given by 
\be 
r_\mu{}^0{}_j = \overset{\scriptscriptstyle(-2)}\Gamma{}^0_{\mu j} = \overset{\scriptscriptstyle(-2)}\Gamma{}^\lambda_{\mu\nu} n_\lambda L_j^\nu \, , 
\ee 
and can also be written in terms of the other fields using the identity $de^{\h{0}} = - \omega^{\h{0}}{}_{\h{j}} \wedge e^{\h{j}}$ \cite{toAppear}. 
The expansion of $\omega^{\h{i}}{}_{\h{j}}$ begins with a $c^2$ term $\omega_\mu{}^{\h{i}}{}_{\h{j}} = c^2 \overset{\scriptscriptstyle(2)}\omega{}_{\mu}{}^{\h{i}}{}_{\h{j}} + \overset{\scriptscriptstyle(0)}\omega{}_{\mu}{}^{\h{i}}{}_{\h{j}} +O(c^{-2})$. 
In particular 
\be 
 \overset{\scriptscriptstyle(2)}\omega{}_{\mu}{}^{\h{i}} {}_{\h{j}}  = \frac{1}{2} n_\mu \sigma_{ij} \, , 
\ee 
and 
\ba 
&&  \overset{\scriptscriptstyle(0)}\omega{}_{\mu} {}^i{}_j  = - L_j^\rho \tilde{D}_{\mu} l^i_\rho + \frac{1}{2} ( A_i \sigma_{\mu j} - A_j \sigma_{\mu i} ) \nn \\ 
&& \hspace{0.5cm} + n_\mu \left[ ( A_i \rho_j - \rho_i A_j ) + \frac{1}{2} (  m^k {}_{j} \sigma_{ki} - m^i{}_k \sigma_{jk}  ) \right] \nn \\ 
&& \hspace{0.5cm} + n_\mu  \frac{1}{2} \beta^{ik} \sigma_{kj} \,   .  
\ea 
Having the explicit form of the spin connection we can write the action for the NC fields using the first order Palatini formalism
\be 
S = \frac{c^4}{8\pi G} \frac{1}{2(d-1)!} \int \epsilon_{\h{a}_1 \dots \h{a}_D} e^{\h{a}_1} \wedge \dots e^{\h{a}_{d-1}} \wedge R^{\h{a}_{d} \h{a}_{d+1}} \, . 
\ee 
Here $R^{\h{a}}{}_{\h{b}} = R^{\h{a}}{}_{\h{b}} (\omega)$ is the Riemann form expressed in terms of the spin connection. Variation of the action with respect to $\omega$ gives the relations written above that express the spin connection in terms of the vielbeins, while variation with respect to the vielbein gives the Einstein field equations. \\ 
\textbf{Expansion of the Ricci tensor and the field equations} \\ 
Using the expression $R^{\h{a}}{}_{\h{b}} = d\omega^{\h{a}}{}_{\h{b}} + \omega^{\h{a}}{}_{\h{c}} \wedge \omega^{\h{c}}{}_{\h{b}}$ for the Riemann form we calculate the expansion of the Ricci tensor. We consider the separate components $R_{\h{0}\h{0}}$, $R_{\h{0}\h{j}}$, $R_{\h{i}\h{j}}$. For the former we obtain 
\ba 
&& \mathcal{R}_{\h{0}\h{0}} =  - \frac{c^4}{4} \sigma_{jk} \sigma_{kj}  + c^2 \left[   \mbox{Tr} (\sigma m \sigma) + \rho_{i} \rho_{i} - \pa_i \left(\rho_i\right) \right. \nn \\ 
&& \left. - \frac{1}{2} \sigma_{ik} \tilde{D} A_{ik}  + \frac{1}{2} \sigma_{ik} \tilde{D} l^k_{i0}  - \frac{1}{4} A_\perp^2 \mbox{Tr}(\sigma^2) \right. \nn \\ 
&& \left. - \frac{1}{4} \mbox{Tr} \left( \sigma \beta \sigma \right)  \right]  
\ea 
plus terms of order $c^0$. For empty space the equations of motion are $R_{\h{a}\h{b}} = 0$, so that at order $c^4$ we obtain $\mbox{Tr} (\sigma^2) = 0$, 
 which implies $\sigma = 0$. Then we recover the result of \cite{van2017torsional} that the geometry is TTNC, where $n \wedge dn = 0$. In the following we will retain $\sigma$ terms and work with a generic torsional geometry. A similar calculation gives the expansion of $R_{\h{0}\h{j}}$:   
\ba 
&&\mathcal{R}_{\h{0}\h{j}} =  c^2 \left[ \frac{1}{2} \pa_i \left(\sigma_{ij}\right) + \sigma_{ik} \tilde{D} l^k_{[ij]} + \sigma_{k[j} \tilde{D} l^k_{i]i}  \right. \nn \\ 
&& \hspace{0.3cm} \left[ + \sigma_{jk} \rho_k - \frac{1}{4} \mbox{Tr} (\sigma^2) A_j + \frac{1}{2} (\sigma \sigma)_{jk} A_k \right] + O(c^0) \, . 
\ea 
The term of order $c^0$ in $\mathcal{R}_{\h{0}\h{j}}$, which is longer, will be displayed in \cite{toAppear}. Finally, the expansion of $\mathcal{R}_{\h{i}\h{j}}$: 
\ba 
&& {R}_{\h{i}\h{j}} = - \frac{c^2}{2} (\sigma\sigma)_{ij} + Ric^{(A)}_{(ij)}  - \rho_i \rho_j + \pa_{(i} \left( \rho_{j)} \right) +  \rho_k \tilde{D}_{(i} l^k_{j)} \nn \\ 
&& + \tilde{D}_{(i} \left( \sigma_{j)k} A_k  \right)   + A_{(i} \sigma_{j)r} \tilde{D}l^k_{kr} + A_{(i} \tilde{D}l^r_{|k| j)} \sigma_{kr}  \nn \\ 
&& + 2 A_{(i} \sigma_{j)k} \rho_k + 2 A_k \sigma_{k(i} \rho_{j)} + A_{(i} d_k (\sigma_{|k|j)}) \nn \\ && - \frac{1}{4} A_i A_j Tr(\sigma^2)  + A_k \sigma_{kl} \sigma_{l(i} A_{j)} - \frac{1}{2} \left( \sigma A\right)_i \left( \sigma A\right)_j  \nn \\ 
&& + (\sigma\sigma m)_{(ij)} +  (\sigma m \sigma)_{(ij)} + r^{0}_{k(i} \sigma_{j)k} - \frac{1}{4} \left( \sigma \beta \sigma \right)_{ij} \nn \\ 
&&  + O(c^{-2}) \, . 
\ea 
We have tested the expansion of the Ricci tensor derived here against the non-trivial metric
\be 
ds^2 = - \left( c T(r) dt + \frac{c x}{\kappa} dr \right)^2 + dr^2 + dx^2 + dy^2 \, , 
\ee 
with $T(r) = a(r) + \frac{b(r)}{c^2}$, $a$ and $b$ being arbitrary functions. We have chosen this metric not for its physical origin, but because it has non-zero values for the torsion components $\rho$ and $\sigma$. We calculated explicitly the vielbein components of its Ricci tensor and found agreement with the tensorial expansion presented in this section in terms of the NC fields. \\ 
\textbf{The generalized Eisenhart-Duval lift metric} \\ 
Consider a scalar, classical particle moving in the metric $g_{\mu\nu}$. Its dynamics is determined by the Lagrangian $H = \frac{1}{2} g^{\mu\nu} p_\mu p_\nu$, where $p_\mu$ is the relativistic momentum. For a particle of mass $m$  the Hamiltonian is conserved, $H = - \frac{m^2 c^2}{2}$. Using the expansion \eqref{eq:inverse_metric_expansion} and solving for $p_0$ we obtain 
\be 
p_0 = A_i p_i + \sqrt{m^2 c^4 + c^2 p_i p_i + (A_i p_i)^2 + \beta^{ij} p_i p_j} \, . 
\ee 
This is the usual relativistic dispersion relation expressed in the NC components, and it is Milne invariant by construction. Traditionally the low energy regime is determined by small values of the spatial momenta, $|p_i| << m c$, however in this setting we will also require a condition on the geometry: $A_k A_k = h^{\mu\nu} A_\mu A_\nu << c^2$ and $|\beta^{ij}|<<c^2$. Then in this limit we can write a non-relativistic dispersion relation 
\be 
p_0 - \left( m c^2 + \frac{1}{2m} p_i p_i + A_i p_i \right) = 0 \, . 
\ee 
This is Milne invariant but is not invariant under the $U(1)$ gauge transformations of the field $A_\mu$, by construction. In the non-relativistic holography literature $A_\mu$ has been associated to the conserved current for the particle number, but in general relativity the particle number is not conserved even for low energies. To write a non-relativistic dispersion relation that is both Milne invariant and $U(1)$ invariant we will modify the former relation by adding  the term $-m \left(A_0 - \frac{1}{2} A_\perp^2\right)$ which, as seen earlier, is Milne invariant and not $U(1)$ invariant. It will be negligible with respect to the rest mass term if $A_0 << c^2$. So we propose the modified dispersion relation 
\be 
p_0 - \frac{1}{2m} p_i p_i - A_i p_i + m \left(A_0 - \frac{1}{2} A_\perp^2 \right) = 0 \, ,  
\ee 
where we have absorbed the rest mass term $mc^2$ into a redefinition of $p_0$. 
According to an established procedure \cite{cariglia2015null,cariglia2018cosmological}, we can introduce a new momentum $p_\xi$, which will be constant and equal to $-m$, associated to a new ignorable coordinate $\xi$, and make this relationship homogeneous and null: 
\be 
p_0 p_\xi + \frac{1}{2} p_i p_i - p_\xi A_i p_i - p_\xi^2 \left(A_0 - \frac{1}{2} A_\perp^2 \right) = 0 \, . 
\ee 
Such null conic is associated to null geodesics of a new Hamiltonian $\mathcal{H} = \frac{1}{2} \h{g}^{MN} p_M p_N$, written in terms of a set of $d+2$ coordinates $x^M = \{ x^\mu, \xi \}$. The metric $\h{g}$ is given by  
\be \label{eq:Eisenhart_generalised}
\h{g}_{MN} dx^M dx^N =  n_\mu dx^\mu \left( d\xi + A_\mu dx^\mu \right) + h_{\mu\nu} dx^\mu dx^\nu \, . 
\ee 
When $dn=0$ this reduces to a standard Eisenhart-Duval lift metric. However for $dn\neq 0$ as we have shown this describes the non-relativistic dynamics in the original spacetime with generic metric $g_{\mu\nu}$. The dynamics now is $U(1)$ invariant as gauge transformations of $A_\mu$ are absorbed by a shift in the variable $\xi$. \\ 
\textbf{Conclusions} \\ 
By embedding torsional NC gravity into General Relativity we have been able to write the equations of motion and the action in generality, without restrictions on the torsion. Thus our approach can be used to study non-relativistic gravity as it interacts with matter, and it can provide a starting point for its quantization. We have explained how the metric \eqref{eq:Eisenhart_generalised}, that has already appeared in the literature, arises from non-relativistic motion in a generic GR metric $g_{\mu\nu}$, thus clarifying its physical meaning and opening new avenues for its study in the context of dynamical systems. Our explicit decomposition of the spin connection can have applications in non-relativistic holography, constructing the action and equations of motion of non-relativistic spinors.

\begin{acknowledgements}
This work is dedicated to Cris and Gaia. I would like to thank Dmitri Sorokin, Gary Gibbons, Peter Horvathy and Christian Duval for useful discussions, and the Theoretical Physics group in Padova for hospitality and financial support during my visit in the early stages of this project. I acknowledge CNPq support from projects (205029/2014-0) and project (303923/2015-6), and FAPEMIG support from the Pesquisador Mineiro project n. PPM-00630-17.  Some calculations used the computer algebra system Mathematica \cite{Mathematica}, in combination with xAct and the xTensor, xCoba packages \cite{xAct}. 
\end{acknowledgements}

\vspace*{-1ex}

%\clearpage   

\end{document}